\documentclass{article}

\usepackage{arxiv}

\usepackage[utf8]{inputenc} 
\usepackage[T1]{fontenc}    
\usepackage{hyperref}       
\usepackage{url}            
\usepackage{booktabs}       
\usepackage{amsfonts}       
\usepackage{nicefrac}       
\usepackage{microtype}      
\usepackage{lipsum}
\usepackage{graphicx}
\usepackage{subfigure}
\graphicspath{ {./images/} }
\usepackage{amsmath}
\usepackage{multicol}
\usepackage{multirow}

\title{PALQA: A Novel Parameterized Position-Aware Lossy Quantum Autoencoder using LSB Control Qubit for Efficient Image Compression}

\author{
Ershadul Haque, Manoranjan Paul,  Faranak Tohidi, Anwaar Ulhaq, Tanmoy Debnath  \\
  School of Computing, Mathematics and Engineering\\
  Charles Sturt University\\
  Bathurst, NSW 2795 \\ Corresponding author email: 
  \texttt{mhaque@csu.edu.au} \\
}
\begin{document}
\maketitle
\begin{abstract}
With the growing interest in quantum computing, quantum image processing technology has become a vital research field due to its versatile applications and ability to outperform classical computing. A quantum autoencoder approach has been used for compression purposes. However, existing autoencoders are limited to small-scale images, and the mechanisms of state compression remain unclear. There is also a need for efficient quantum autoencoders using standard representation approaches and for studying parameterized position-aware control qubits and their corresponding quality measurement metrics. This work introduces a novel parameterized position-aware lossy quantum autoencoder (PALQA) circuit that utilizes the least significant bit control qubit for image compression. The PALQA circuit employs a transformed coefficient block-based modified state connection approach to efficiently compress images at various resolutions. The method leverages compression opportunities in the state-label connection by applying position-aware least significant control qubit. Compared to JPEG and other enhanced quantum representation-based quantum autoencoders, the PALQA circuit demonstrates superior performance in terms of the number of gates required and PSNR metrics.  
\end{abstract}

\section{Introduction}
\label{sec:intro}
In the era of data science, image data is widely used in various fields, such as social media, the medical field, and beyond. In quantum technology, images play a crucial role in visualization, data analysis, communication, and innovation across various quantum applications, including quantum computing, quantum communication, quantum imaging, and quantum machine learning. Visual representations of quantum information enable researchers, developers, and users to better understand and leverage quantum principles in diverse technological domains. With increasing image data, day by day, it becomes a new challenge in quantum computing, such as storage, transmission, and restoration\cite{wang2024quantum}. Image compression involves reducing redundant information while preserving important information\cite{wang2024quantum}. The compression of digital images is becoming a challenging issue with the advent of quantum technology due to varying processor architectures. The emerging field of quantum information processing (QIP) has opened up a new era in information processing.

Moreover, quantum machine learning (QML) is a growing field of interest with various applications, including image classification, image denoising, sensing, anomaly detection, transfer learning, and feature learning\cite{biamonte2017quantum,rebentrost2014quantum, lloyd2014quantum,havlivcek2019supervised,wu2023efficient}. Furthermore, an autoencoder is an artificial neural network used for unsupervised learning, dimensionality reduction, feature learning, and data reconstruction. The main idea behind an autoencoder is to learn a compressed and efficient representation (encoding) of the input data, which can then be used to reconstruct the original data (decoding)\cite{hecht1989neurocomputing}. 
Quantum technology offers several advantages over classical computation, including faster processing, enhanced security, and higher accuracy, due to the utilization of three fundamental properties of quantum mechanics: parallelism, superposition, and entanglement \cite{grover1996,nielsen2001quantum}. 

\begin{figure}
    \centering
    \subfigure[Classical autoencoder]
    {
     \includegraphics[width=0.45\textwidth, height=8cm]{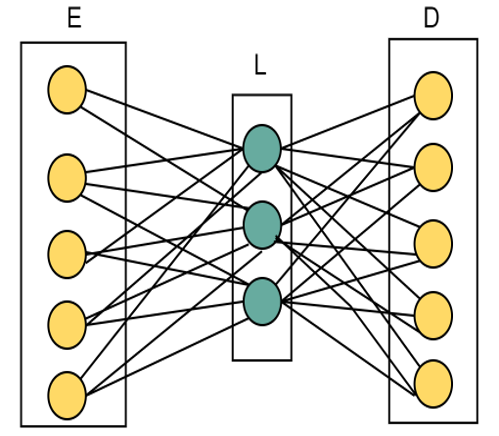}
     \label{fig_cae}
    }
    \subfigure[Quantum autoencoder]
    {
    \includegraphics[width=0.45\textwidth, height=8cm ]{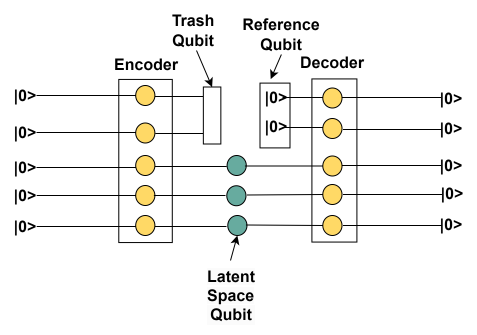}
    \label{fig_QAE}
    }
    \caption{autoencoder architecture. (a) Architecture of the classical autoencoder (E) encodes $n$ dimensional input data into the lower dimensional of the latent space (L) and the decoder (D) decodes original information from the lower dimension of the latent space. (b) Quantum autoencoder architecture where the encoder(E) takes an $m$ number of qubits state and maps into a lower number latent space(L) qubit and then the decoder (D) reconstructs the image into a $m$ dimensional state.}
    \label{fig_QAE_CAE}
\end{figure}

Quantum autoencoders can compress quantum states into a smaller number of qubits and decompress image information from the reduced qubit data using swap qubit information. The success of quantum machine learning approaches mainly depends on the efficient loading of classical data into quantum systems\cite{wu2023efficient}. This is crucial for reducing limited and costly resources in quantum computation and communication. Figure \ref{fig_cae} illustrates a classical autoencoder, which encodes the information into a reduced form and transmits it through a transmission channel to the decoder. The decoder then reconstructs the original information from this reduced data. In the case of lossy compression, some original information cannot be fully recovered due to the inherent loss in the compression process.

A quantum autoencoder (QAE) concept is shown in Figure.\ref{fig_QAE} and has been proposed by Romero et al. 2017 \cite{romero2017quantum}. It compresses the information by sending it to the trash qubit while encoding information. For reconstruction, the reference qubit means trash qubit information is used to preserve the discarded state information. It is a lossless compression based on the FRQI approach. How and where the compression happened is still unclear because it feeds the same information from the trash qubit to the decoder side. How the pixel and its corresponding position connect is still another gap. Moreover, when dealing with many inputs, it experiences limited fidelity bound \cite{cao2021noise}. Also, the reconstruction measurement quality of the image was not performed regarding compression capability relating to specific connected qubit information. However, quantum information processing using quantum circuits is a common practice for transmitting image data \cite{giovannetti2008quantum,bharti2022noisy}. 
The classical information is converted into quantum data using a parameterized encoding circuit and processed using specific quantum machine learning tasks\cite {wu2024quantum}—for example, ref. \cite{grant2018hierarchical} employed an autoencoder circuit with a four qubits network to translate public Irish image datasets \cite{fisher1988iris} into a quantum circuit and its corresponding state. The existing QAE stored information in the quantum circuit approaches is unable to compress quantum image information directly related to the pixel and image. Therefore, considering the direct compression issue of pixel and state(position), further study is needed to find an efficient quantum autoencoder circuit. The quantum autoencoder circuit is considered a generalized QAE network.  

 Quantum and classical technologies are not entirely dissimilar. In quantum technology, qubits propagate from left to right within a long wire system, whereas classical computers use binary values of zero (0) and one (1). A qubit is represented as a column matrix formed by the binary values 0 and 1. In quantum circuits, existing quantum autoencoders face challenges in compressing stored qubit information due to the limitations of their circuit connection architecture \cite{wu2024quantum}.
On the other hand, the parameterized quantum circuit plays a crucial role in many classical quantum algorithms and is highly efficient for classification tasks. Selecting an efficient circuit is another critical and challenging aspect due to its dependency on the encoding process of various image pixel values, including state connections. This is because it determines the required number of qubits that connect pixel values with their corresponding state connections. A parameterized circuit diagram offers one of the best solutions for directly working with quantum operational gates, allowing the compression of the state connection for each pixel value. Two different models are typically used for encoder and decoder circuits of quantum autoencoder \cite {thesis}. The first model is the FRQI approach \cite{le2011flexible}, which encodes image colour and state connection using a control rotational gate. Due to the probabilistic outcome, it is very challenging to reconstruct the original
pixel value, including its corresponding state, from the output.
Although, it is limited to the $2\times2$ image size. Also, how pixel intensity and its state are encoded using a single control of the rotational gate is still a question.  

Zhang et al. \cite{zhang2013neqr}proposed a NEQR approach for representing image data in the quantum photonic circuit. The issue with this method is that it is limited to small images. Also, it encodes each one in the pixel value separately, including its position. As a result, more operational gates are required to complete the circuit connection to represent the whole image pixel information, including the state. Moreover, reconstructed image quality measurement is another gap in the NEQR approach. It does not relate to the compression issue. Later, Flip (2021) demonstrated that the NEQR approach exhibits better performance compared to FRQI-based QAE circuits\cite{thesis}. In \cite{thesis}, $2\times2$ image is reconstructed from the outcome of probabilistic mapping value. How the position is encoded in the quantum circuit is still not clear. A modified version of the NEQR approach proposed in \cite{haque2023enhancing} discards the identified gate for circuit implementation since identity gates do not affect output. As a result, it becomes very efficient for representing images and compression further via the binary truncation approach. Haque et al. (2023) proposed a ZSCNEQR circuit that represents transform coefficient values using quantization and DCT approach\cite{haque2023efficient}.    

In this paper, a novel position-aware parameterized lossy quantum (PALQA) based on LSB control qubit quantum autoencoder is proposed to improve the existing quantum autoencoder approach for compression of grey-scale images using a quantization, qubit swap, and lossy preparation approach including other quantum processing approaches. The new model uses the quantized transform coefficient associated with the state connection of the qubits sequence after block-wise division and position-aware LSB qubits as a swap qubit. The position representing qubits stores the positional information of the transform coefficient value of the grey-scale image for the first time. Eight qubits store the quantized transformed coefficient information to represent the quantized transform value. Moreover, an improved formula has also been proposed to achieve lossless positional value recovery. For $8\times8$ division of block position, it needs $(log_28+1)$ and $(log_28+1)$ number of qubits to recover the position of the quantized transformed coefficient. 
Furthermore, it established a condition by swapping a fixed position qubit information to achieve a highly positional aware quantum autoencoder with higher latent space, such as 14. Another great potential of the proposed approach is the ability to compress and reconstruct images with a higher compression ratio and provide a better PSNR value. Furthermore, BPP and PSNR are used as performance metrics to evaluate compression capability, as they provide a deterministic measurement approach. In addition, specific qubit information is computed for the first time to determine the significant qubits that carry most of the quantum information. Because of the above additional features, the newly proposed quantum autoencoder circuit is more flexible and very efficient for image compression as a fundamental quantum autoencoder approach compared to all other existing quantum autoencoders. 

The remaining sections of the paper are organized as follows: Section \ref{L_R} outlines related works, Section \ref{P_M} discusses the proposed approach, Section \ref{R_D} analyzes the computational results and provides a discussion, and Section \ref{CC} concludes the findings of this work.

\section{Related Works}\label{L_R}
Ref \cite{chiribella2015universal} proposed two unitary gates for reducing the dimension of the input gates and another two unitary gates to reconstruct an original number of gates. It addresses the question of the possibility of super-replication of the gates, which is considered totally from an image perspective. A cloud computing-based quantum autoencoder was proposed in \cite{zhu2023quantum} for reducing the number of qubit resources for communication. It addresses the problem of more quantum resources between client and server regarding communication purposes. However, it considers single qubits, whereas most generalized quantum circuits are multi-qubits.  

A higher fidelity-based QAE was proposed in \cite{bravo2021quantum} for enhancing data loader efficiency. It presents the enhanced fea-
ture quantum autoencoder (EF-QAE) and demonstrates its
performance through simulations. There is still a gap in how compression happens through latent space. In addition, the standard representation model used was not mentioned clearly. Moreover, it is limited to two qubits  latent space only, which is very tiny to represent a size image. Also, how only two qubits will represent the colour and its corresponding state is not appropriately investigated. Furthermore, another issue is which representation model has been used to incorporate qubit connections, such as connecting state qubits and colour qubits. Also, how compression happens in the ground states must be mentioned clearly. Besides, rate-distortion performance metrics such as PSNR versus the required number of gates per pixel were not measured based on the reconstruction image quality. 

A single photon-based quantum autoencoder has been proposed to compress quantum resources in \cite{pepper2019experimental}. It uses a classical optimization approach to utilize quantum resources such as state. It can learn the fixed structure of the data. The reconstruction of the state, including color, is still a gap. Also, how it applies in the case of image compression is not mentioned clearly. In \cite{buhrman2001quantum}, a swap test  QAE approach has been proposed to identify fingerprints. It compares two states of the fingerprint to identify the original one. Ref~\cite{achache2020denoising} proposed a quantum autoencoder circuit for quantum secret sharing (QSS). It addresses the gap in the speed required to run thousands of qubits. A high-rank mixed-state quantum autoencoder was proposed in \cite{cao2021noise}. Another gap is how compression happens with image color, including state connection.

In~\cite{wang2024quantum}, a hybrid quantum autoencoder was proposed to compress image data into lower latent space. The cost function and gradient were measured to evaluate model efficiency. There is still a gap in how quantum information is mapped into latent space incorporating qubit connection. A machine learning quantum autoencoder approach was proposed in \cite{huang2020realization} that compresses two-qubit states into one qubit. It provides a lossless compression scheme. It is limited to a single latent space qubit. Ref~\cite{bondarenko2020quantum} proposed an unsupervised quantum autoencoder for denoising cluster state. In \cite{du2021exploring}, an effective QAE  learning protocol was proposed to address the reduction technique of the high dimensional space into lower dimensional latent space. It can calculate corresponding eigenvalues with a lower rank of state property to measure error. 

Tom et al. (2020) reported a QAE circuit to correct Green-berger-Horne-Zeilinger (GHZ) states using depolarizing and bit flip channels \cite{achache2020denoising}. Gong et al. (2024) explored a quantum neural network using a quantum variational circuit angle and amplitude encoding. Due to angle and amplitude encoding, it generates probability outcomes. It is used for image classification purposes, not for compression. In 2022, Zhang et al. proposed a high-dimensional quantum autoencoder for teleportation of compressed quantum data \cite{zhang2022resource}. It uses integrated photonic in a compress-teleport-decompress manner. The key concept of this approach is to compress dimensional by throwing redundant information and recovering the original state using a teleportation mechanism. Park et al. (2023) described a variational one-class classifier to address the one-class classification issue using a semi-supervised fully-parameterized quantum autoencoder \cite{park2023variational}. It uses a probabilistic approach using a control rotational gate. 

To enhance predicted output accuracy toward mitigating error, a fixed-scale quantum circuit including one hot encoding approach was implemented in \cite{li2024quantum}. Quantum circuit consists of a sequence of gates and the nature of timing. Anand et al.(2022) proposed classically immutable circuits for the compression of the quantum information \cite{anand2022quantum}. It demonstrated various quantum state compression using a low representation of quantum data. There is still a gap in how the image is encoded for both color and corresponding states. In quantum machine learning, a high dimensional function estimator is another approach to avoid over-fitting issues via parameterized quantum circuit~\cite{chen2021expressibility}. It described error scalability using ansatz saturation and related between generation and expressibility using circuit depth.  

\begin{figure}[htbp]
\centerline{\includegraphics[width=0.45\linewidth, height=0.35\linewidth]{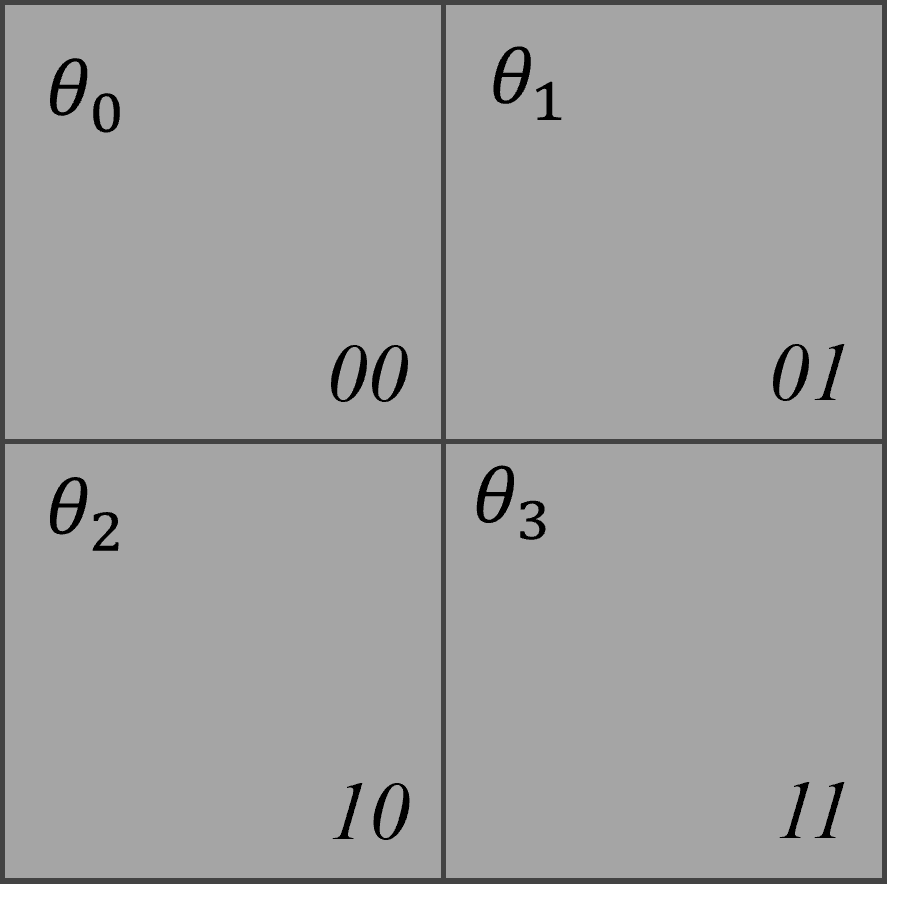}}
\caption{A $2\times2$ FRQI quantum image}
\label{FRQI_A}
\end{figure}

Tabi et al. (2022)  proposed a hybrid-classical quantum neural network to mitigate the noise of the quantum processor using the noisy channel. A factorized quantum measurement circuit has been presented to map the probability distribution of the state via quasi-stochastic and local unitary matrices~\cite{carrasquilla2021probabilistic}. A statical floating-point lossy quantum algorithm presented in \cite{yoon2022lossy} incorporating binary learning approach. Besides, the bias correlation approach is also applied to minimize error via inexact reconstructed data. A tree structure hybrid amplitude encoding approach has been proposed to classify image data complying with angle encoding~\cite{gong2024quantum}. It demonstrated the width and height adjustment of the image related to the quantum circuit resources. Liu et al.(2024) designed a quantum algorithm to compress classical information using hidden sub-group concept~\cite{liu2024information}. A flexible representation of the quantum image (FRQI) approach is generally used in the encoder and decoder section for quantum autoencoder architecture. It is proposed by Le, which encodes the pixel value using a control rotational gate \cite{le2011flexible}. Figure ~\ref{FRQI_A} represents $2\times2$ FRQI image, including its states. 
It can be expressed as, \\

\begin{math}
|I_{FRQI}\rangle=\frac{1}{2}[\left(cos\theta_{0}|0\rangle +sin\theta_{0}|1\rangle \right) \otimes |00\rangle 
+\left(cos\theta_{1}|0\rangle+sin\theta_{1}|1\rangle \right)\otimes|01\rangle +  \left(cos\theta_{2}|0\rangle+sin\theta_{2}|1\rangle \right)\otimes|10\rangle+ \left(cos\theta_{3}|0\rangle+sin\theta_{3}|1\rangle \right)\otimes|11\rangle ]\nonumber
\end{math}

\begin{figure}[t!]
\centerline{\includegraphics[width=0.8\linewidth,height=4cm]{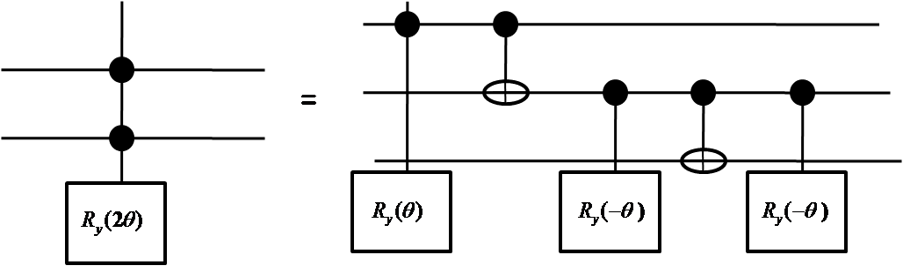}}
\caption{An FRQI circuit for representing $|I_{FRQI}\rangle$ image}
\label{FRQI_Circuit}
\end{figure}

Figure ~\ref{FRQI_Circuit} depicts the circuit for representing   $|I_{FRQI}\rangle$ image~\cite{le2011flexible}. 

Where, $R_y(2\theta)$ indicates standard rotation metric and is given as,\\
$R_y(2\theta)$=
$\begin{pmatrix}
\centering
  cos\theta_i & -sin\theta_i \\ 
  sin\theta_i & cos\theta_i
\end{pmatrix}$ 
\\
Generally, it represents image pixel value using a control rotational gate. The control rotation gate can be implemented using standard rotation and c-not gate. It is unable to represent pixel-wise grey-scale complex operational value due to the use of a single qubit. It represents color information as an angle and position in a qubit sequence. Generally, it uses a control rotating gate and stores pixel value in the Bloch sphere. Due to the probabilistic measurement approach, it is very challenging to reconstruct the original image. Most of the autoencoder is based on FRQI architecture. 

\begin{figure*}[htbp]
\centerline{\includegraphics[width=\linewidth,height=8cm]{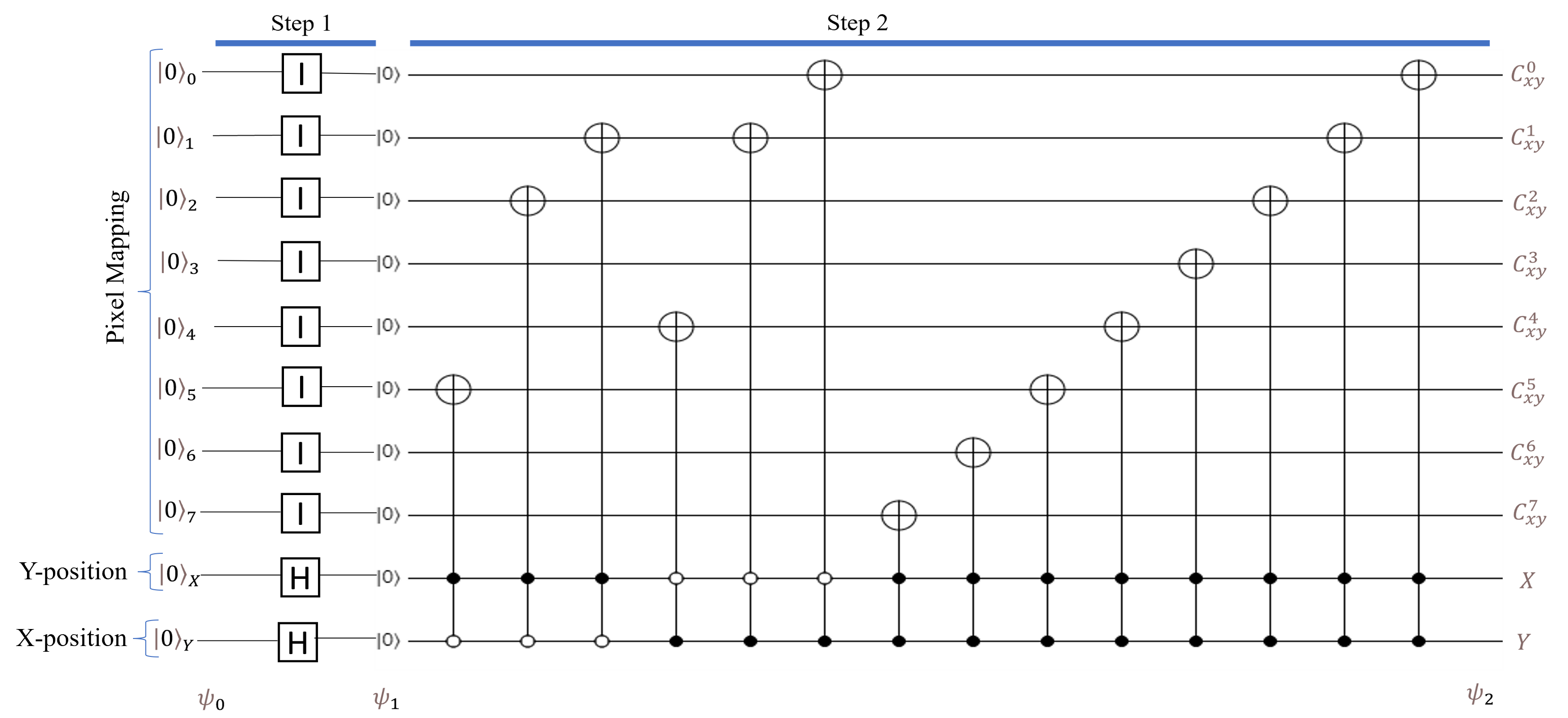}}
\caption{An NEQR circuit for pixel values representation}
\label{NEQR_Circuit}
\end{figure*}

After that, Zhang et al. (2013) proposed a novel enhanced quantum representation(NEQR) approach \cite{zhang2013neqr} that represents the grey-scale pixel value. It converts pixel values into a binary system, where only a frequent number of ones are considered when mapping the pixel value. In a quantum system, pixels and positions representing qubits are generally used to map quantum circuits. For example, an image containing pixel values are $0 (Y=0, X=0), 100 (Y=0, X=1), 200 (Y=1, X=0),$ and $255(Y=1, X=1)$. The pixel value's quantum representation is known as the NEQR approach and is expressed as $|I_{NEQR}\rangle$.    

\begin{flalign}
\resizebox{0.9\hsize}{!}{$
|I_{NEQR}\rangle=\frac{1}{2}[|0\rangle\otimes|00\rangle 
+|100\rangle\otimes|01\rangle+|200\rangle\otimes|10\rangle+|255\rangle \otimes|11\rangle]\nonumber
$}
\end{flalign}
Figure~\ref{NEQR_Circuit} shows the circuit diagram of NEQR approach for $|I_{NEQR}\rangle$ image.   

\begin{figure*}[htbp]
\centerline{\includegraphics[width=\linewidth,height=8cm]{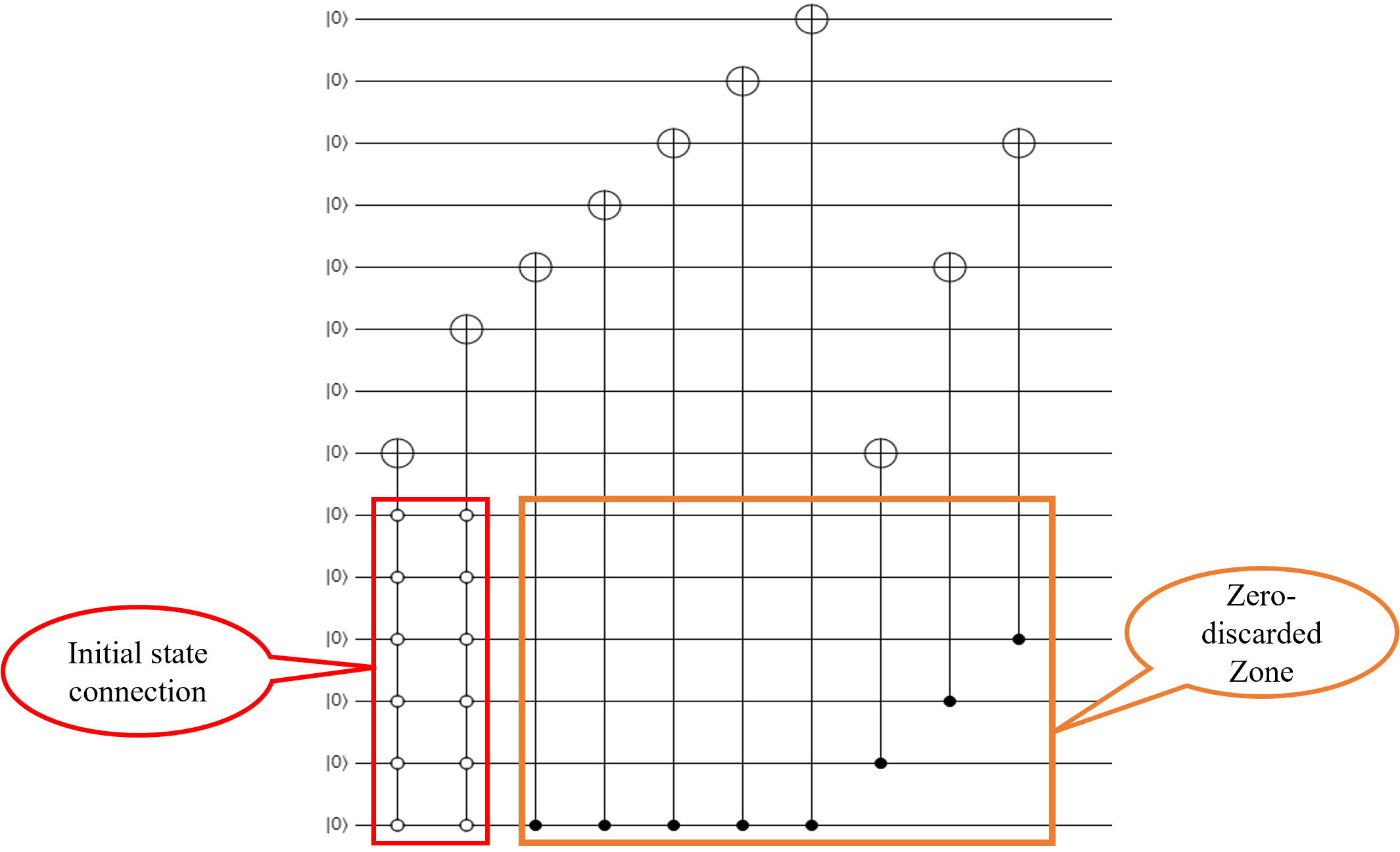}}
\caption{A NZ-NEQR circuit for pixel values representation where an initial connection (marked as red) and zero-discarded zone (marked as orange). Zero is discarded because the identity gate has no control over the c-not gate.}
\label{fig_nzneqr_diagram}
\end{figure*}

It addresses FRQI issues and provides two qubits: one for color representation and another for the state of the grey-scale image's pixel value. An improved architecture of the NEQR circuit has been proposed in \cite{haque2023enhancing}. It addresses the higher resource requirement of the quantum gates to connect the pixel value. It uses the transfer coefficient value and its corresponding position to implement the NZ-NEQR circuit, which requires several connections to complete the image connection. For example, Figure~\ref{fig_nzneqr_diagram}, shows the circuit diagram of the NZ-NEQR approach for pixel values of $5(X = 0, Y = 0), 248(X = 1, Y = 0), 1(X = 2, Y = 0), 8(X = 4, Y = 0)$ and $32(X=0, Y=1)$ respectively.

\begin{figure*}[!t]
\centerline{\includegraphics[width=\linewidth,height=7cm]{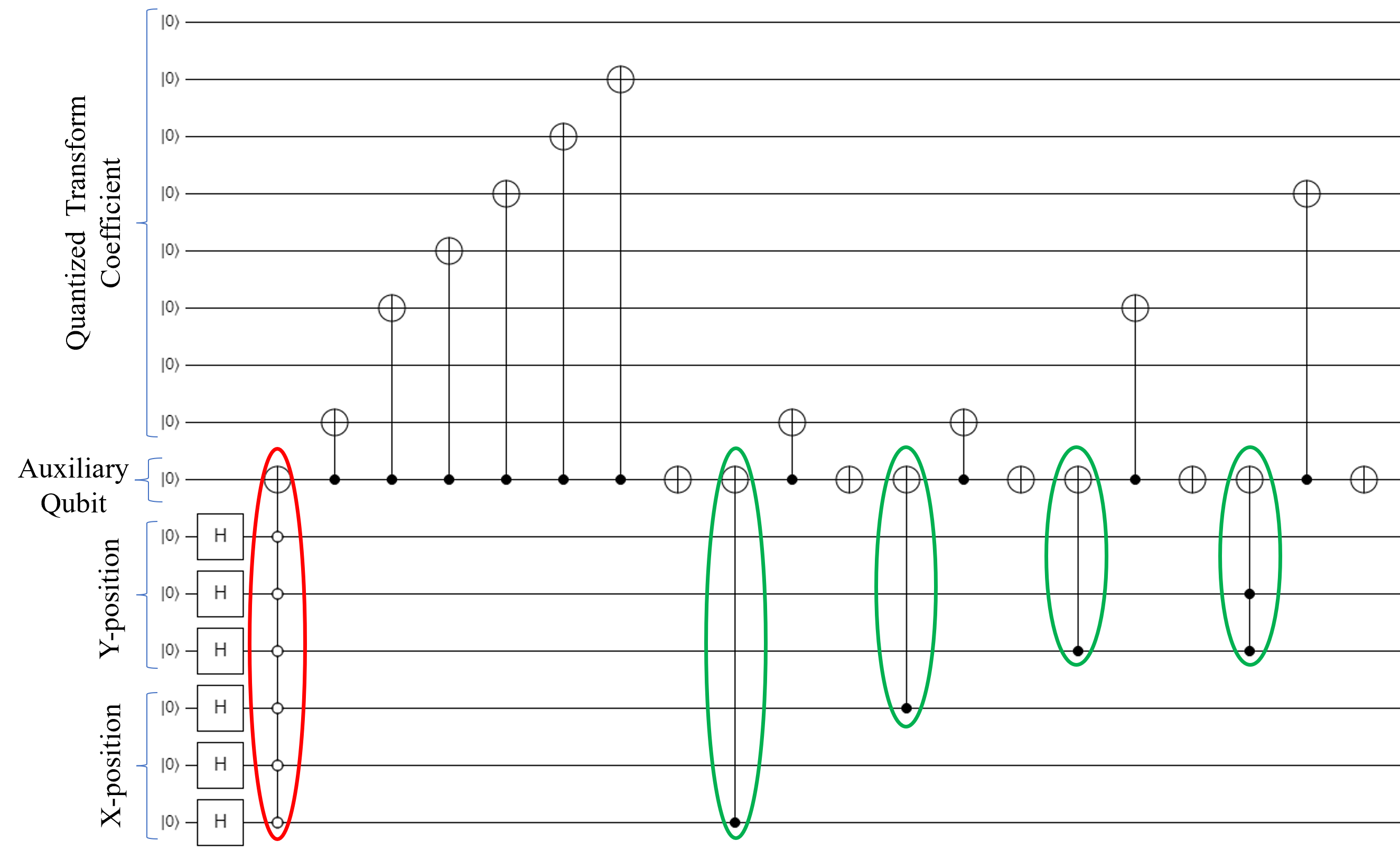}}
\caption{Quantized transform coefficient representation based on ZSCNEQR circuit. It includes an initial connection (marked in red) and a zero-discarded zone (marked in green)}
\label{fig_proposed_SCMNEQR_diagram}
\end{figure*}

In ~\cite{haque2023efficient}, a modified version of the SCMNEQR (state connection modification novel enhanced quantum representation) approach  \cite{haqueblock} is known as ZSCNEQR (zero-discarded state connection novel enhance quantum representation) approach that minimizes the required number of gate connections. It is more efficient than the SCMNEQR approach because it provides better rate-distortion performance metrics via block-wise division of the transfer coefficient value. For example, Figure~\ref{fig_proposed_SCMNEQR_diagram} shows ZSCNEQR circuit diagram for quantized transfer coefficient,  $125(X=0, Y=0), 1(X=1, Y=0), 1(X=4, Y=0), 4(X=0, Y=1),$ and $16(Y=3, X=0)$ values.

Flip et al. 2021 demonstrated that NEQR quantum autoencoders perform better than the FRQI approach for $2\times2$ image metric~\cite{thesis}. Although it provides a better result than the FRQI approach, there are still many challenges with the NEQR-based quantum autoencoder, such as limited to $2\times2$ image. Moreover, the state label encoding system is not mentioned correctly. Every time, the state position relates to 1's connection with the transfer coefficient value, which is challenging. Performance is not assessed to measure its capacity for compression and reconstruction. Due to the probabilistic outcome, it does not mention how compression happens regarding quantum information. Besides, reconstruction quality measures were not investigated. Moreover, how it relates to the standard NEQR approach was not investigated. In addition, how medium and higher-resolution images are encoded using NEQR autoencoder is another gap. To address the above issues of the existing NEQR QAE, further study needs to improve the architecture. 

\section{Proposed methodology}\label{P_M}
\begin{figure}
\centering
    {
    \includegraphics[width=.48\textwidth, height=5.5cm]{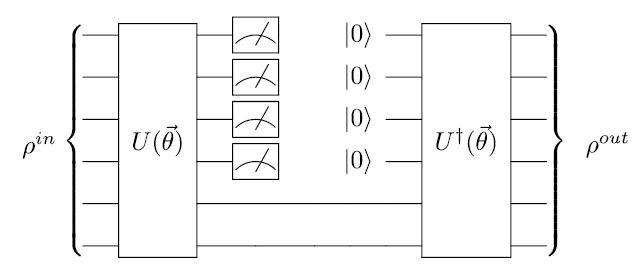}
    }
\caption{Traditional 6-2-6 quantum autoencoder circuit for ground state compression of the Hubbard and Hamilton model, where the unitary quantum encoder encodes 6 qubits into 2 qubit state and unitary decoder attempting to recover 6 qubit input onto the 6 qubit output.}
\label{Traditional_approach}
\end{figure}

\begin{figure*}
\centering
    {
    \includegraphics[width=\textwidth, height=8cm ]{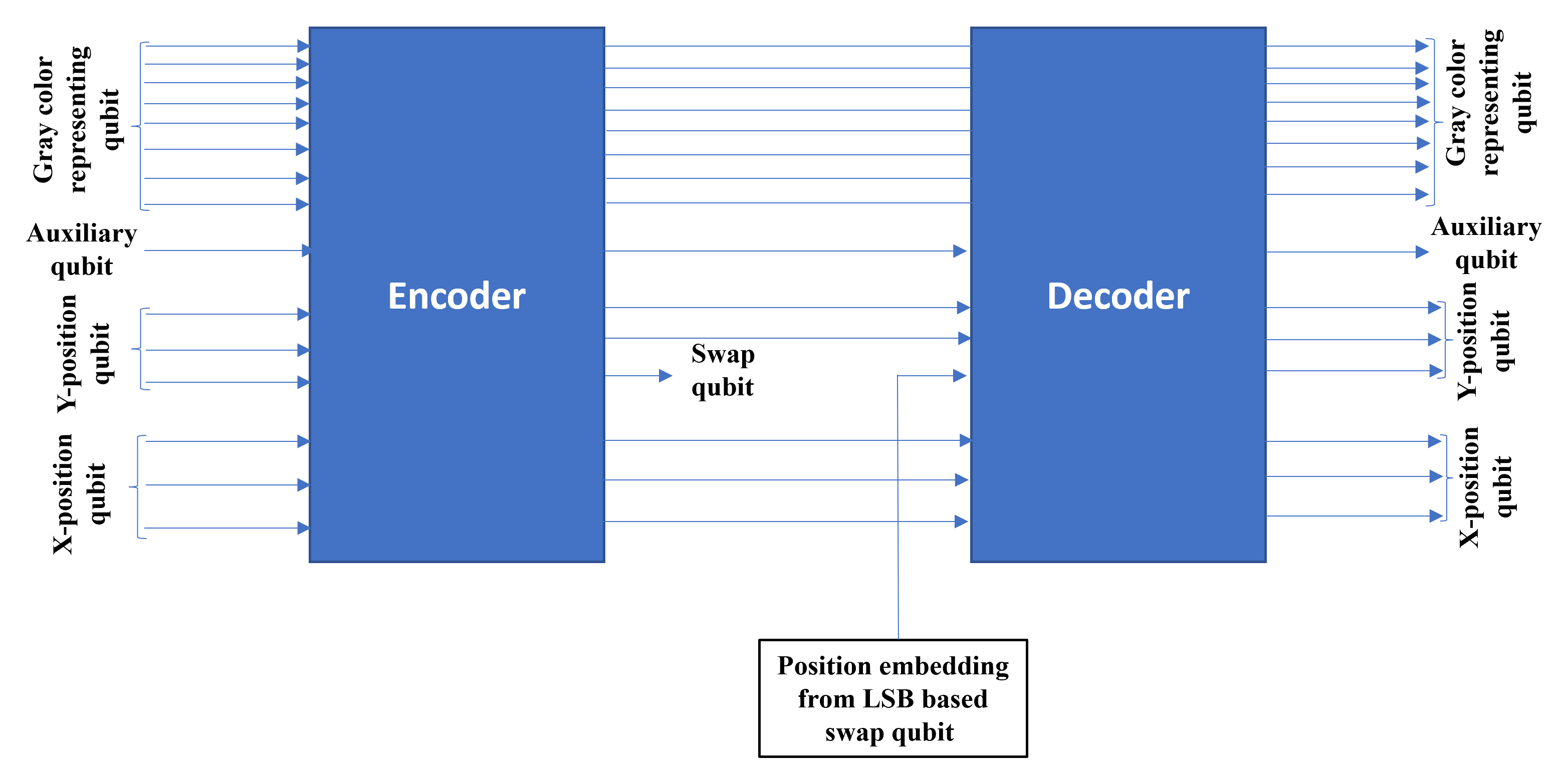}
        \label{Prposed_approach}
    }
\caption{Proposed 14-13-14 quantum autoencoder circuit, where encoder encodes color and position separately and decoder decodes the original color and position from latent space representing qubit.}
\label{proposed_architecture}
\end{figure*}

This section describes the proposed QAE architecture compared to the traditional QAE approach. Figure~\ref{Prposed_approach} shows the proposed approach architecture compared to the traditional autoencoder Figure.\ref{Traditional_approach}. Ramora et al. (2017) proposed the very first quantum autoencoder to compress the ground state of the molecular Hamiltonians and Hubbard model \cite{romero2017quantum}. It does not relate to how and in what way the image could be compressed. 

In the proposed system, the modified ZSCNEQR circuit is used to map the transform coefficient value and its corresponding state position because of its potential to match transfer coefficient values with lower requirements of gate connection while maintaining a higher PSNR value. After that, the LSB positional qubit is considered a swap qubit at the encoder side. Then, information on the swap qubit based on the state position qubit is replaced with the trash qubit directly. Only 1's positional value in the swap qubit is transmitted to the receiver. A new function generates the discarded information at the decoder side to concentrate on the same positional value of discarded 1's information from the LSB positional qubit.   
Quirk simulation tool has been used to design the modified ZSCNEQR circuit \cite{b17}. Matlab is also used to compute quantum resources. A $2^a \times 2^b$ image size is considered for compression purposes. 
In the designing of the modified ZSCNEQR approach, the steps involved are: \\
Step 1: Prepare a quantized DCT coefficient value. Then, calculate the required number of qubits to represent the quantized transfer coefficient and its corresponding position to represent the qubit. It required $(q+n+1)$ number of qubits, where $q$ is the required qubit for quantized transfer coefficient value and $n=\log_2(S)$, $S$ is the number of blocks that X- and Y-positional value of the quantized transfer coefficient. For $8\times8$ block position of the quantized transfer coefficient, 4 ($log_2X+1$) number of qubits are required to connect the maximum value of the X-positional block. The same number of qubits is also required to connect the Y-positional value.\\
Step 2: The below equation represents the initial state of the quantized transfer coefficient, 
 \begin{equation}
   |\Psi_0\rangle={\vert0\rangle}^{\otimes(q+2n+1)}
 \end{equation}
 
$q+1$ identity gate and $n$ number of Hadamard gate are required to implement the modified ZSCNEQR approach. Since a grayscale image is considered to validate our proposed approach, 8 qubits are needed to map the quantized transfer coefficient value, and one additional qubit is also required as an auxiliary qubit to map the auxiliary qubit information. On the other hand, for state mapping, 8 qubits are necessary to map the position representing qubit information. 
The encoding procedure of the quantized transfer coefficient can be represented as, \\
\begin{equation}
U=I^{\otimes{q+1}}\otimes H^{\otimes{n}}
\end{equation}

In addition, the initial state can converted into an intermediate state using the $U$ transformation factor, 

\begin{equation}
\Psi_1=U(|\Psi_0\rangle)=(I|0\rangle)^{\otimes{q+1}}\otimes (H|0\rangle)^{\otimes{n}}
\end{equation}

Where $\Psi_0$ is the initial state and $\psi_1$ is the intermediate state. 
The $U_2$ converts the intermediate state of the quantized transfer coefficient into the final state. The below equation describes the final preparation procedure. 

\begin{flalign}
\Psi_2&=U_2(|\Psi_1\rangle) \\  
&=\frac{1}{2^n} \sum_{i=1}^{n-1}\sum_{j=1}^{n-1}\,\left(|C_{YX}\rangle (|Y_{o}X_{o}\rangle \right)
+\frac{1}{2^n}|C_{Y_zX_z}\rangle|Y_{z}X_{z}\rangle) \nonumber
\end{flalign}

$|C_{YX}\rangle$ is the quantized transform coefficient value of 1's only whose position is $Y_{o}X_{o}$. On the other hand, $|C_{Y_zX_z}\rangle$ represents the transform coefficient value of the zero ($|Y_{z}X_{z}\rangle)$) position. Then, the LSB qubit of the X-position is extracted using the below formula, 
\begin{flalign}
X_p=X_{LSB} \nonumber
\end{flalign}

$X_p$ extract the LSB qubit information from the X-position representing qubit. The below equation describes $U_2$ quantum operator, \\
\begin{equation}
U_2=\prod_{X=0,....,2^n-1}\prod_{Y=0,....,2^n-1}\, \left(U_{Y_{z}X_{z}}+U_{Y_{o}(X_elsb)}\right)
\end{equation}
Where $X_elsb$ indicates the required number of 1's information excluding  LSB information of the X-positon. 
The required number of connections for transmitting the state connection information is expressed,\\

 \begin{equation}
          B_{state}= (log_2(X_0)+1+log_2(Y_0)+1-X_{LSB}+X_{LSB,ones})\otimes{Tc_{nz}}
 \end{equation}

 $Tc_{nz}$ represents the non-zero quantized transfer coefficient values. $X_{LSB}$ indicate LSB qubit information including both 0 and 1.  Whereas, $X_{LSB, 1's}$ represents the frequent number of 1's in the LSB qubit.$B_{state}$ indicates the required number of total gates for connection to the state value of the quantized transfer coefficient. The total number of required connections at the transmitter end is, \\
 \begin{equation}
{B_{total}}=q_{ones}+B_{state}+B_{sign}+B_{auxilary}+B_{gpp}
 \end{equation}

Where $q_{ones}$ indicates the required number of gates for connecting quantized transfer coefficient value to complete the autoencoder circuit. $B_{sign}$ is the required number of sign bits for transmitting the quantized transform coefficient sign value. $B_{auxilary}$ is the required number of gates from auxiliary qubits to connect its corresponding transfer coefficient value with its position. $B_{gpp}$ is the gates per pixel(gpp) locate the block total block positioning error.    
 
The number of required connections indicates the required number of gates to implement the proposed network. Each connection indicates the gates and the total number of required gates indicates the total number of bits to implement the proposed circuit. Therefore, the required number of gpp is expressed as, 

\begin{equation}
     {gpp}=\frac{B_{total}}{I_s} 
 \end{equation}

\begin{figure*}
\centering
    \subfigure [Encoder architecture]
    {
        \includegraphics[width=0.9\textwidth, height=8cm]{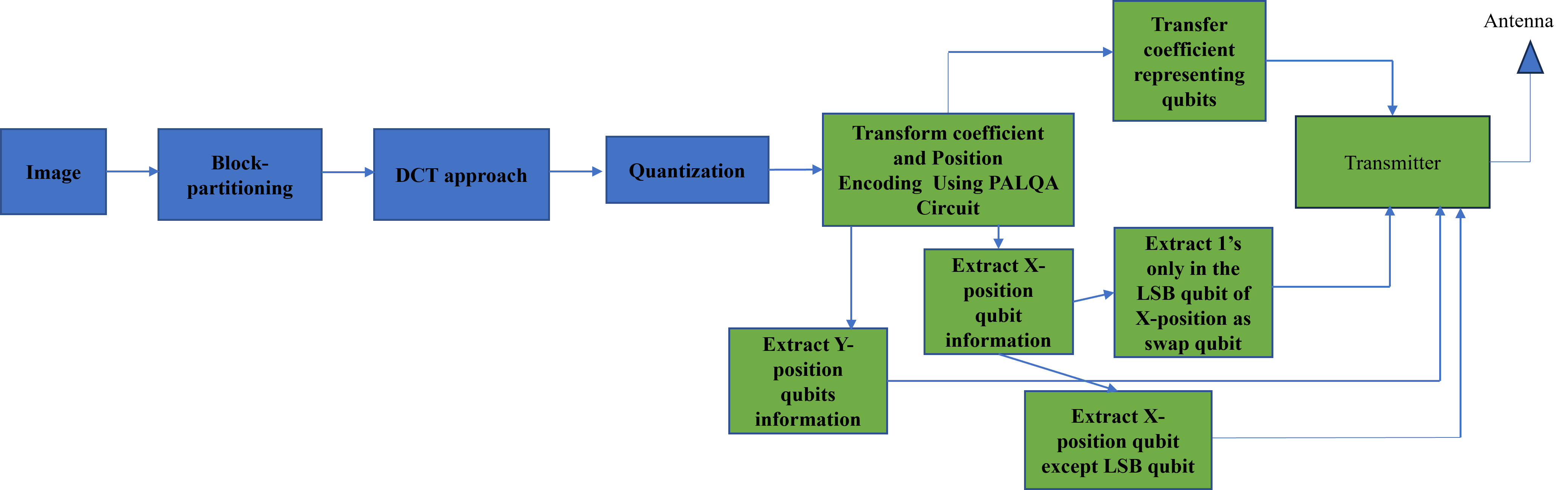}
        \label{Encoder_architecture_pro}
    }
        \subfigure [Decoder architecture]
    {
        \includegraphics[width=0.9\textwidth, height=8cm]{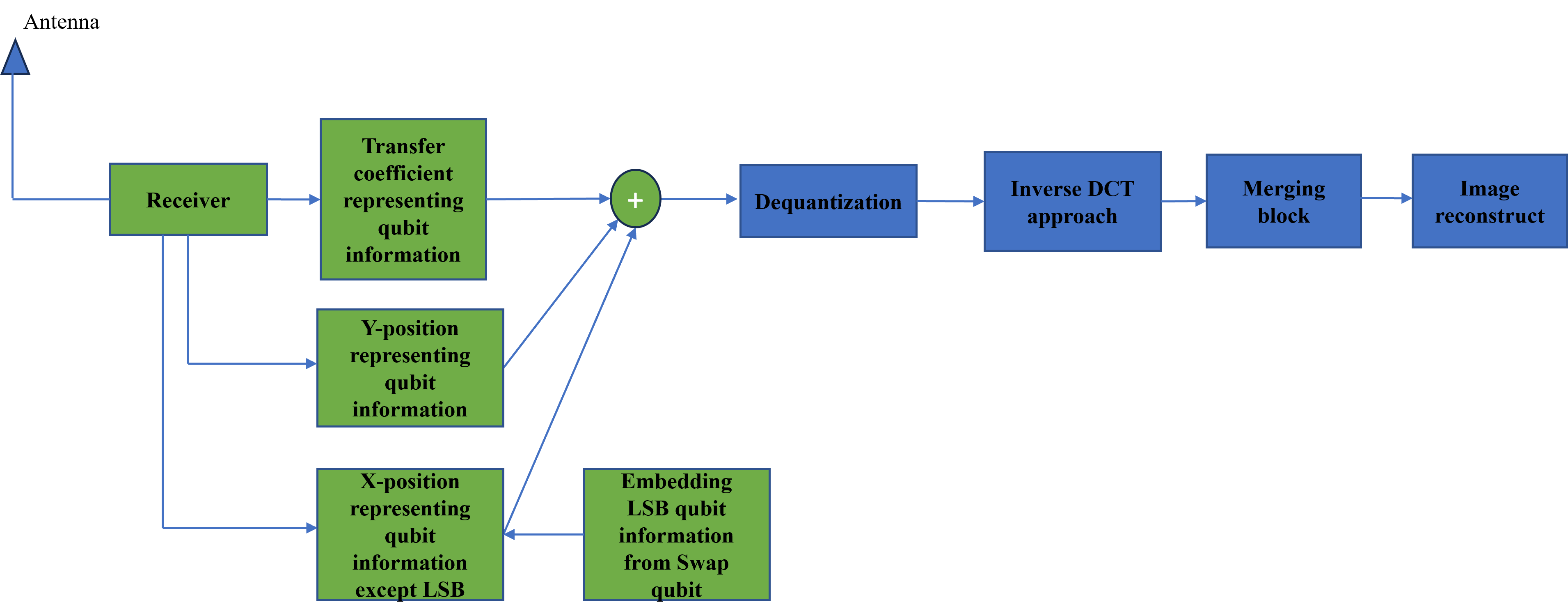}
        \label{Decoder_architecture_pro}
    }
\caption{a) the architecture of the proposed encoder architecture, in which blue color represents the classical pre-processing and green color shows the quantum processing parts. b) decoder architecture of the proposed approach, where the receiver receives the signal and processes further using quantum and classical processing techniques.}
\label{proposed_architecture_encoder}
\end{figure*}

Figure~\ref{Encoder_architecture_pro} shows the proposed encoder architecture, consisting of the classical and quantum parts. The blue block diagram shows the classical part, while the green shows the quantum computation. An image is partitioned into $8\time8$ block. After that, apply the DCT transformation approach on the $8\time8$ block. Quantize the transform DCT coefficient using various scalar quantization factors. Encode the quantized transform coefficient value using the proposed PALQA circuit. PALQA circuit is the modified version of the ZSCNEQR approach. The modification is done in the auxiliary qubit connection. Then, the X- and Y-positions representing qubit information of the quantized transform coefficient are extracted. Then, the Y-position representing qubit information is sent directly to the transmitter. The transmitter transmits that information via a transmitting antenna. Besides, extracting X-positional qubit information from the quantized transfer coefficient value and LSB qubit information is considered a swap qubit. Preserve the frequent number of 1's value in the LSB qubit information of the X-position as swap qubit information, including its positional value. Subsequently, the frequency of 1's is then sent to the transmitting antenna via the transmitter block.            
Figure~\ref{Decoder_architecture_pro} shows the decoder architecture of the proposed approach. It consists of the receiver, several processing blocks for the quantized transfer coefficient, its corresponding position, adder, dequantization, inverse DCT, merging block, and image reconstruction. The receiver receives the transmitted signal and sends it to the further processing of the quantized transfer coefficient, including its positional qubit information. The transfer coefficient representing block reform the quantized coefficient values, including sign operation. The Y-position representing qubit information is directly sent to the adder operation. For lossless recovery of the Y-position, 4-qubits are used to encode the Y-positional information because of the maximum positional value.\par
On the other hand, for recovery of the X-position's LSB qubit information, use a new function that generates LSB qubit information from the swap qubit but requires less information. Specifically, a logical concept is used to generate the swap qubit information. In the logic, two conditions are considered:\\

\textbf{Condition 1:} 
\begin{itemize}
\item Generate the same dimensional of the LSB matrix if all the elements in the LSB    qubit connection are zero.  
\end{itemize}
\textbf{Condition 2:} 
\begin{itemize}
\item if all the elements in the LSB qubit connection are not zero, read and preserve the number of 1's including its position.   
\end{itemize}

After that, feed the LSB qubit positional information into the X-position representing qubit information except the LSB block. This block concatenates LSB positional qubit information with the existing qubit information. As a result, every information in the X-position representing qubit is recovered. The adder generates a new quantized transform mapping the recovered X- and Y-positional qubit information. Consequently, the dequantization block de-quantizes new quantized transfer coefficient values and feeds to the inverse DCT approach. The block is merged hereafter, and the image is reconstructed using the image reconstruction block. In such a way, zero in the LSB qubit can be discarded, contributing to the X-position's compression. Since the swap qubit information is fully retrieved, that is why the system is lossless.  

\subsection{Modified ZSCNEQR circuit and LSB qubit swap}

\begin{figure*}
\centering
    {
        \includegraphics[width=\textwidth, height=6cm ]{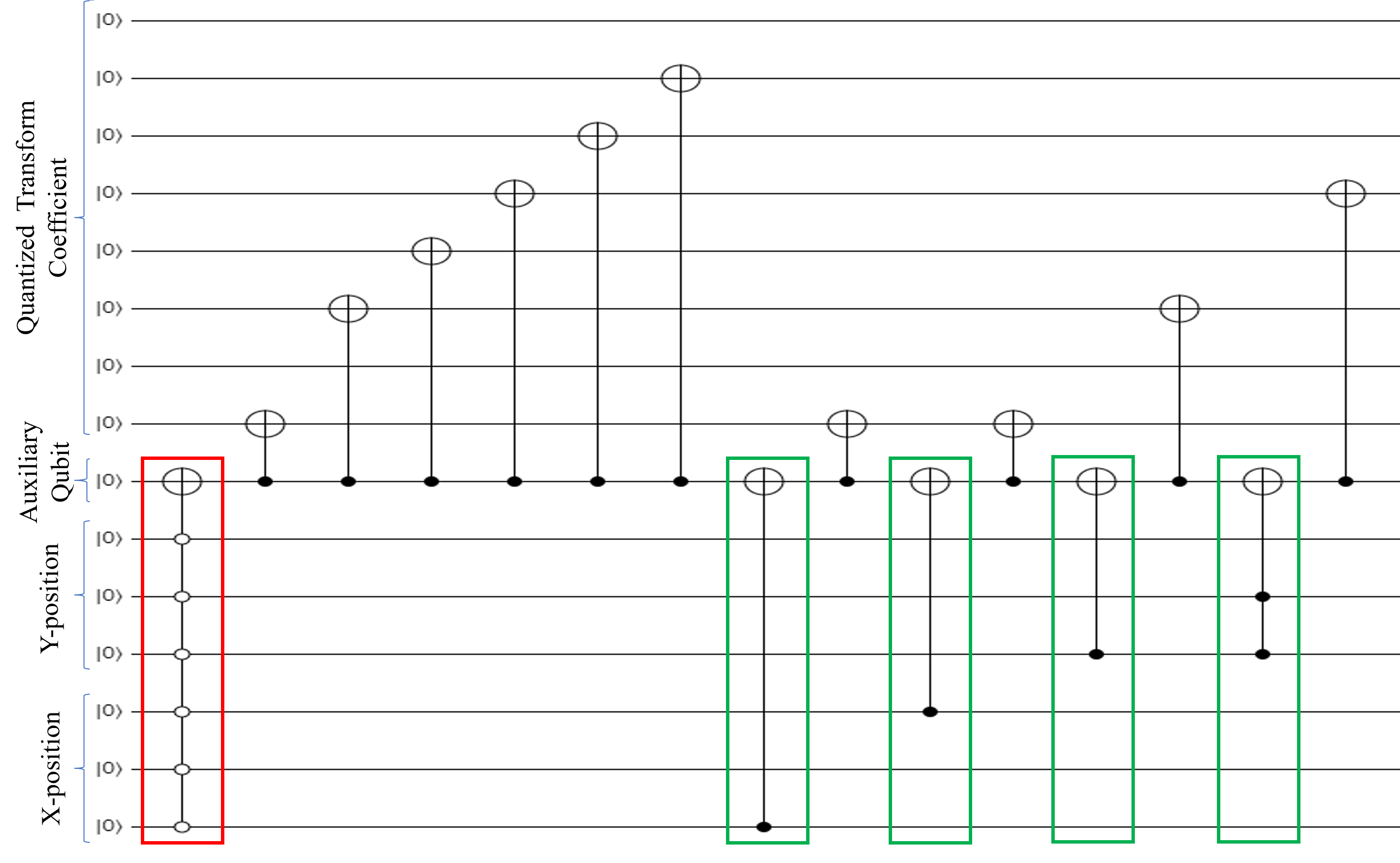}
    }
\caption{A modified ZSCNEQR circuit for mapping quantum autoencoder information.}
\label{modified_ZSCNEQR_approach}
\end{figure*}

Figure~\ref{modified_ZSCNEQR_approach} shows the modified ZSCNEQR circuit for mapping the quantum information. The red mark in Figure~ \ref{modified_ZSCNEQR_approach} shows the initial state connection of the state representing qubit that connects the state position qubit connection to the auxiliary qubit to locate the quantized transform coefficient values. The difference between this circuit and the ZSCNEQR approach is that a reset gate is used after connecting each quantized transfer coefficient value. On the other hand, the green zone shows the identity gate mean (zero) omitting zone in the state representing zone. The auxiliary qubit connects the quantized transfer coefficient value to its corresponding state position. The X-position and Y-position of the quantized transfer coefficient values are connected using the X-position and Y-position representing qubits. The bottom of the qubit is the LSB qubit, while the top upper qubit is the most significant bit (MSB) qubit.        

\begin{figure}
\centering
    {
        \includegraphics[width=0.45\textwidth, height=4cm ]{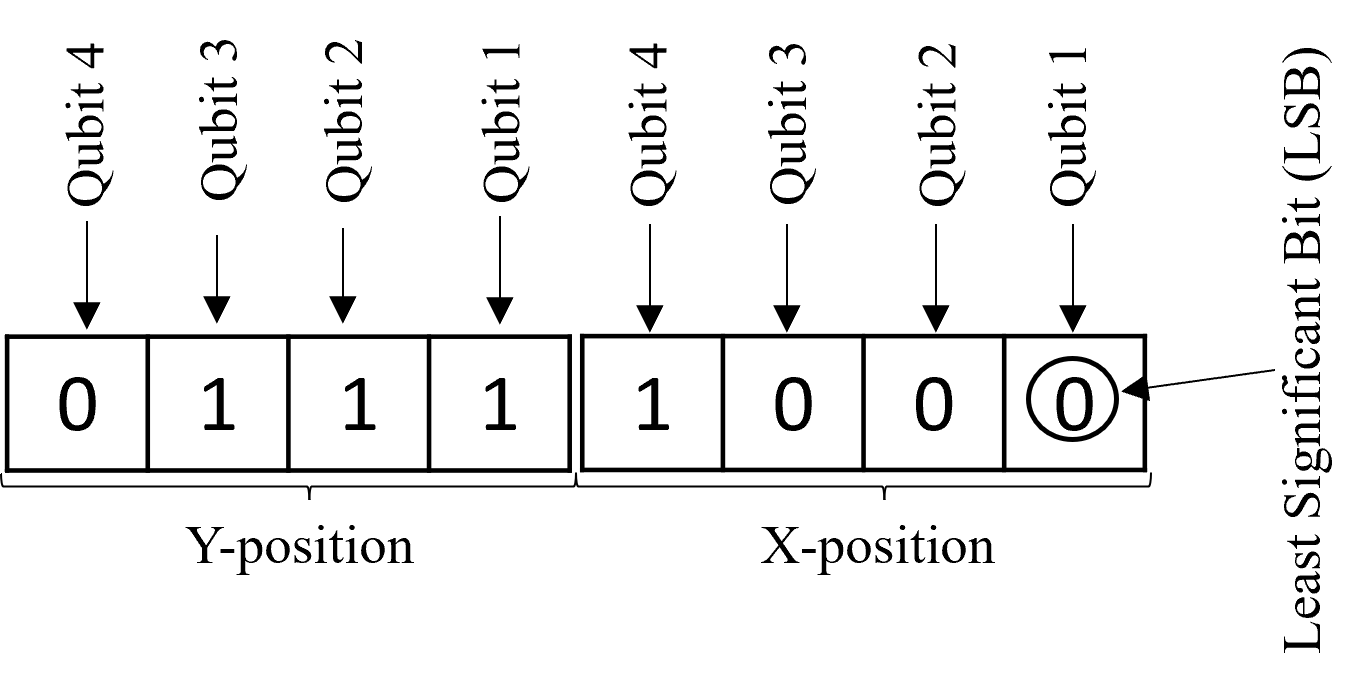}
    }
\caption{LSB qubit of the state value of the X-positional qubit.}
\label{LSB_qubit_connection}
\end{figure}

In image steganography, LSB substituting the pixel value is a common technique, replacing the original information of the LSB bit. The least significant bits(LSB) contain the least amount of information\cite{yu2023adaptive}. Therefore, changing LSB has a negligible effect on the overall image. In quantum computing, position recovery is very sensitive to loss of positional information. As a result, LSB qubit information is replaced by swap qubit. Then, using a proposed function generation means protocol, the LSB positional qubit information of the transfer coefficient is regenerated. After that, LSB positional qubit information is concatenated with the function generator. The function generator regenerates the same information discarded from the encoder end. In such a way, the LSB qubit positional information swap technique becomes lossless for position recovery. For example, Figure~\ref{LSB_qubit_connection} shows the LSB qubit of the X-position of the quantized transfer coefficient approach. LSB qubit contains the LSB information used to connect the LSB qubit connection to the auxiliary qubit to locate the LSB state positional value. Only the number of 1s is considered with its respective positional value.           

\section{Result and discussion}
\label{R_D}
In this section, the computation result related to the proposed approach has been analyzed for cameraman$(224\times224)$, baboons$(512\times512)$, scenery $(512\times512)$, peppers$(512\times512)$, airport$(1024\times1024)$, and building $(512\times512)$ grayscale images \cite{bb17,bb18} using various quantization factors and preprocessing approaches. Besides, there are two different approaches for the compression of images: one is Jpeg, and another one is NZ-NEQR-based quantum autoencoder, also simulated for the considered images. The same quantization factors are considered, and the result shows the RD performance value to investigate each method's compression capability. After that, the proposed approach's computational results are compared with those of the classical compression approach, such as Jpeg and NZ-NEQR-based quantum autoencoder. 

\begin{figure*}
\centering
    \subfigure [cameraman image]
    {
        \includegraphics[width=0.45\textwidth, height=5cm]{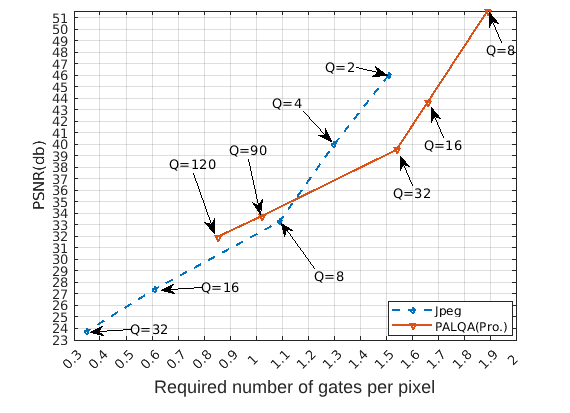}
        \label{gray_channel_cameraman}
    }
    \subfigure[baboon image]
    {
        \includegraphics[width=0.45\textwidth, height=5cm ]{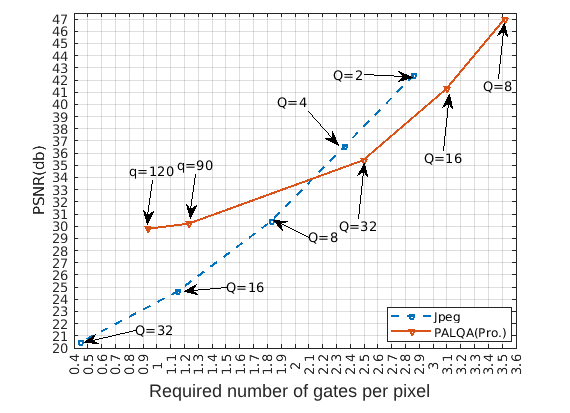}
        \label{gray__channel_baboonsr}
    }
    \subfigure[scenery image]
    {
        \includegraphics[width=0.45\textwidth,height=5cm ]{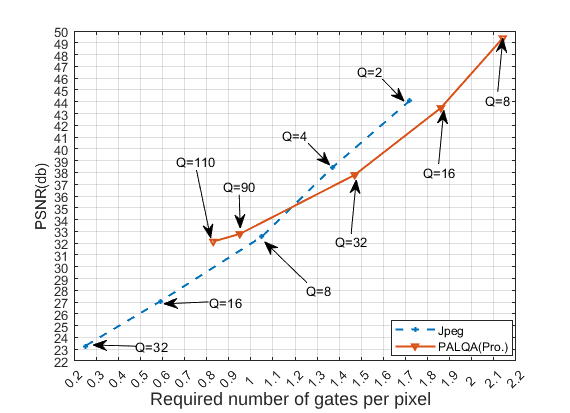}
        \label{gray__channel_scenery}
    }
    \subfigure[pepper image]
    {
        \includegraphics[width=0.45\textwidth,height=5cm ]{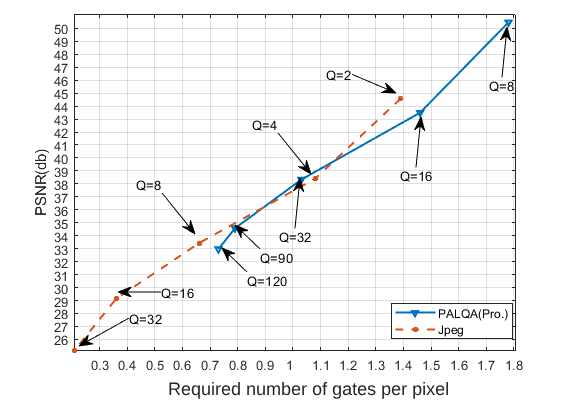}
        \label{gray_channel_peppers}
    }
    \subfigure[airport]
    {
        \includegraphics[width=0.45\textwidth,height=5cm ]{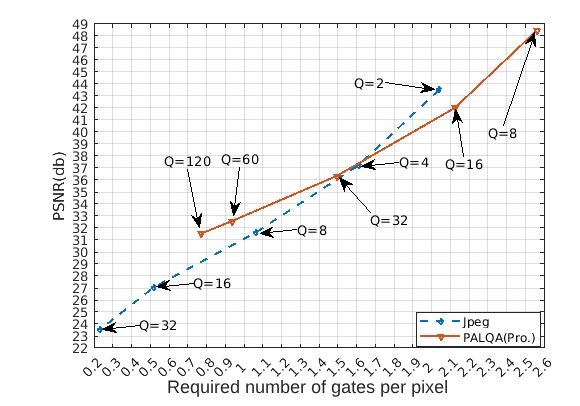}
        \label{gray_channel_airport}
    }
     \subfigure[building image]
    {
    \includegraphics[width=0.45\textwidth,height=5cm]{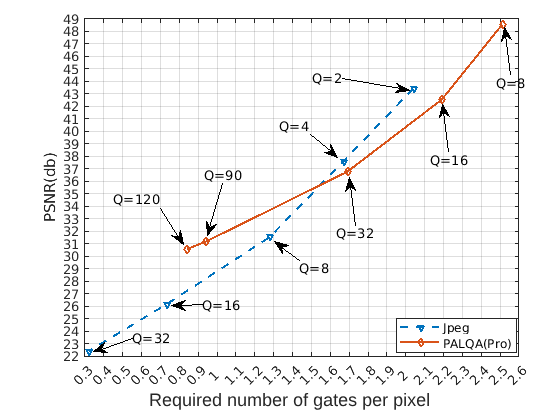}
    \label{gray_channel_building}
    }
\caption{RD performance of the proposed approach compared to Jpeg approach for the gray channel of benchmark images, where qubit number 4(LSB qubit) of the X-position is considered as trash qubit. The green color shows the quantum processing whereas the blue color represents the classical.}
\label{PSNR_comparison_gray_image_jpeg_pro}
\end{figure*}

Figure~\ref{PSNR_comparison_gray_image_jpeg_pro} shows the RD performance of the proposed approach compared to the traditional Jpeg compression approach for considered images using various quantization factors such as 2,4,8, 16, 32, 60, and 90. Due to the use of the quantization factor, the proposed system becomes a lossy compression approach. For the Jpeg approach, 2,4,8,16, and 32 quantization approaches are used to compress the considered images to verify the proposed result compared to the Jpeg approach. For the proposed approach, 8, 16, 32,60, 70,90,110, and 120 quantization factors are used to find the best comparison region compared to the Jpeg and NZ-NEQR-based QAE. For the NZ-NEQR approach, 8,16,32, 70, and 90 quantization factors are used for result comparison. Two performance metrics, such as gpp and PSNR values, are computed for each quantization factor to demonstrate the visual effect of compression of the algorithm. \par

Figure~\ref{gray_channel_cameraman} depicts the computational result of the proposed approach in the case of a grayscale cameraman image compared to the JPEG approach using various quantization factors. For the Jpeg approach, 2,4,8,16 and 32 quantization factors are used to compute corresponding gpp and PSNR values. The X-axis and Y-axis show the gpp PSNR values. For Q=2, Jpeg provides 46.02 db by transmitting a 1.50 gpp value. On the other hand, the PALQA approach provides 51.6db PSNR by transmitting a 1.89 gpp value using an eight-quantization factor. In the case of the gpp value, PALQA draws 0.38 additional gpp value compared to the Jpeg bpp value (1.51) but provides a higher difference in PSNR, such as 5.58db. In the case of the first quantization value, the proposed approach performs superior to the Jpeg approach. For the Q=4 quantization factor, Jpeg requires a 1.30 bpp value to transmit a 39.98db PSNR value.
On the other hand, PALQA draws 43.66db PSNR and 1.66 gpp value using eight quantization factors, block-divisioning, and LSB qubit information swap technique. To gain a 3.99db value, the PALQA requires an additional 0.3 gpp value compared to the Jpeg approach. For the second quantum factor, PALQA performs better than the Jpeg approach for the cameraman approach. For the third quantization value, 8 and 32 factors are used for the Jpeg and PALQA approaches, respectively. Jpeg draws 1.09 bpp and 33.27db PSNR values. Conversely, PALQA  has achieved 1.54 gpp and 39.93db PSNR values. A higher PSNR value indicates the higher similarity of the compressed image to the original image. Identifying the difference between a pre-compressed image and a post-compressed image is very difficult. For 16 quantization factors, Jpeg requires 0.61 bpp to achieve a 27.41db PSNR value.\par
On the other hand, the proposed approach requires 1.02 gpp to achieve a 33.75db PSNR value. PALQA has gained approximately 6.34db more than the Jpeg approach, while it needs an additional 0.41 gpp value. Jpeg uses 16 quantization factors as the final quantization factor. Jpeg provides 23.75db PSNR and 0.31 bpp values for 16 quantization factors. On the contrary, PALQA requires a 0.85 gpp value to achieve a 31.95db PSNR value. In the case of PSNR value, PALQA draws an 8.2db  difference than Jpeg. PALQA needs 0.54 additional gpp value than Jpeg in . The range of the PSNR is 0 to 100. When the PSNR value exceeds 30, the human eyes can not perceive the difference between compressed and original images.   
It is noticed that the proposed approach provides a higher similarity of the compressed image to the original than the Jpeg approach for the cameraman image.\par 

The proposed PALQA approach RD performance metrics of the grey channel of the baboon image is shown in Figure~\ref{gray__channel_baboonsr} compared to the Jpeg approach using 
2,4,8,16,32,90, and 120 quantization factor. Jpeg uses 2,4,8,16,32 quantization factor while PALQA uses 8,16,32,90,120. For Q=8, PALQA draws 46.91 db PSNR  and 3.51 gpp values, while Jpeg attains 42.4db PSNR and 2.86 bpp values using two quantization factors. PALQA gains 7.5db PSNR by sacrificing 0.65 gpp value in addition to the Jpeg approach. JPEG attained 36.58db PSNR for four quantization factors by transmitting a 2.36 bpp value. Oppositely, PALQA achieves a 41.27db PSNR value by transmitting a 3.09 gpp value using 16 quantization factors. 4.69db PSNR has been gained by the PALQA approach compared to the Jpeg approach while sacrificing 0.73 gpp value. For Q=32, PALQA provides 35.40db PSNR and 2.49 gpp values. For the Q=8 quantization factor, Jpeg executes 30.42db and 1.83 bpp values. PALQA gained nearly 4.98db higher PSNR value while it requires an extra 0.66 gpp value compared to Jpeg. PALQA produces a 30.20db PSNR value by transmitting a 1.20 gpp value using a 90 quantization factor. On the other hand, Jpeg provides 24.64db PSNR and 1.15 bpp values. PALQA gains 5.56db higher PSNR while it requires a further 0.5 gpp than the Jpeg approach. Q=32 and 120 quantization are considered final quantization factors for the Jpeg and PALQA approach. Jpeg requires a 0.20 bpp value to achieve a 20.48db PSNR value. In contrast, PALQA provides a 29.78 db value while it requires 0.93 gpp. It gains a 9.3db higher PSNR value while it requires an extra 0.73 gpp value compared to the Jpeg approach. \par

Figure~\ref{gray__channel_scenery} shows the required amount of the gpp versus PSNR values for the scenery image compared to the Jpeg approach using 2,4,8,16,32,90, and 110.
Jpeg provides 44.1db PSNR and 1.72 bpp values for initial quantization, while PALQA produces 49.38db PSNR and 2.14 gpp values. Regarding the PSNR difference, PALQA has gained 5.28db but lost 0.42 gpp. Still, it has attained a noticeable PSNR value compared to JPEG. For the second quantization, 4 and 16 quantizations are used to compute corresponding performance matrices for both Jpeg and PALQA approaches. PALQA attained 43.47db PSNR by transmitting a 1.86 gpp value. On the other hand, Jpeg draws 38.43db PSNR and 1.37 bpp value. PALQA has gained a 5.04db PSNR value by considering the supplement 0.49 gpp value. 32 and 8 quantization are considered the third quantization factor for the PALQA and Jpeg approach. PALQA requires a 1.47 gpp value to attain a 37.78db value, while Jpeg requires a 1.05 bpp value to achieve 32.55db PSNR. PALQA has gained 5.23db PSNR, but it needs an additional 0.42 gpp value compared to the Jpeg approach. For quantization, Q=90, PALQA achieves nearly 32.78db PSNR and 0.95 gpp values. However, Jpeg provides 27.06db and 0.59 bpp values using 16 quantization factors. 
The PALQA approach has gained 5.72db compared to the Jpeg approach. In the meantime, it requires approximately 0.36 gpp value to gain 0.72db, which is reasonably remarkable. 
For Q=32, Jpeg generates 23.27db PSNR by transmitting a 0.25 bpp transmission coefficient value. Alternatively, PALQA has achieved 32.14db PSNR by transmitting a 0.83 gpp value. The PALQA approach has gained a 9db  PSNR difference, while 0.58 additional gpp is required compared to the Jpeg approach.            
\par
The RD performance of the proposed approach for the pepper image is shown in Figure~ \ref{gray_channel_peppers}  compared to the Jpeg approach using 2,4,8,16,32,90, and 120 quantization factors. 2,4,8,16, and 32 quantization factors are used for the Jpeg approach, while 8,16,32,90, and 120 are used for the PALQA approach. Using quantization factors 2 and 8, both Jpeg and PALQA provide 44.58db and 50.49db by transmitting 1.39 bpp and 1.78 gpp values, respectively. PALQA has attained nearly 5.91db higher PSNR value while it requires approximately 0.39 gpp value compared to Jpeg. Considering the gained value of the PSNR, the proposed approach shows its robustness. Using four quantization factors, Jpeg draws a 38.4db PSNR and 1.08 bpp. Conversely, PALQA achieved a 43.49db PSNR value by transmitting a 1.46 gpp value, including transfer coefficient and positional information, using 16 quantization factors. A notable difference is found in the PSNR value, which is 5.09, gained by the PALQA approach compared to the Jpeg approach. In the case of the gpp requirement, it requires approximately an extra 0.38 gpp. For Q=8, Jpeg produces a 33.43db value by transmitting nearly 0.66 bpp.\par
On the other hand, PALQA provides a 38.4db PSNR signal, which requires transmitting around 1.08 gpp value. While PALQA has gained around 4.97db more, it still needs 0.42 additional gpp value than the Jpeg approach. For Q=90, PALQA has achieved nearly 33.43db PSNR and 0.66 gpp values. However, Jpeg provides nearly 29.14db PSNR and 0.36 bpp value using 16 quantization factor. Compared to PSNR and gpp value, it has gained a 4.26db PSNR value and lost 0.36 gpp values compared to the Jpeg approach. The PSNR gained value by the proposed approach is much higher than the gpp value. For Q=32, Jpeg requires roughly 0.21 bpp to perform a 25.18db PSNR signal. Contrariwise, when sending approximately 32.97db PSNR signal, PALQA must transmit nearly 0.73 gpp signal. It achieves approximately 7.79db PSNR value compared to the Jpeg approach. In the case of gpp, it requires around 0.51 ancillary gpp values compared to Jpeg.  \par                  
Figure~\ref{gray_channel_airport} shows the RD performance of the proposed approach of the airport image compared to the Jpeg approach using 2,4,8,16,32,60, and 120 quantization factors. For the Jpeg approach, 2,4,8,16 and 32 quantization factors are used to demonstrate RD performance metric while 8,16,32,90, and 120 are used for the proposed approach. For Q=2, Jpeg provides approximately 43.5db PSNR value by transmitting a 2.04 bpp value. On the other hand, PALQA produces 48.43db PSNR and 2.56 gpp using eight quantization factors. Compared to the PSNR value, PALQA has gained 4.9db compared to the Jpeg approach. Compared with gpp, PALQA requires about 0.5 extra gpp value than Jpeg. For Q=4 and 16, Jpeg and PALQA draw 36.44db and 42.01db PSNR values. Besides, the required gpp values are 1.49 and 2.12. PALQA requires an additional 0.43 gpp value to gain a 5.57db PSNR value. For Q=8, Jpeg provides a 31.59db PSNR value while it needs to transmit a 1.06 bpp. Contrary, PALQA requires a 1.4 gpp value to transmit a 36.33db PSNR signal using a 32 quantization factor. However, PLAQA needs around 0.36 further gpp value, but it has also obtained 4.74 db more PSNR value than the Jpeg approach. Jpeg produces 27.07db PSNR and 0.52 bpp value using the Q=16 quantization factor. On the contrary, PALQA achieves a 32.57db PSNR signal and  0.93 gpp values using the Q=60 quantization factor. The comparison result shows that PALQA achieved a 5.5db PSNR value by considering an extra 0.41 gpp value. \par

Figure~\ref{gray_channel_building} depicts the RD computational result of the proposed approach using 8,16,32,90 and 120 quantization factors, positional LSB encoding, and several pre-processing approaches for building images compared to the Jpeg approach. It has mentioned that Jpeg uses 2,4,8,16, and 32 quantization factors. For Q=2, Jpeg provides a 43.43db PSN signal while it needs to transmit approximately 2.04 bpp value. In the meantime, PALQA achieves 48.53db PNSR and 2.51 gpp values using eight quantization factors. The comparison result shows that PALQA gains a 5.10db PSNR value but needs an additional 0.47 gpp value compared to the Jpeg approach. For Q=4, Jpeg produces a 37.6db PSNR value while it needs to send around 1.67 bpp value.
On the other hand, PALQA attains a 42.63db PSNR value by sending approximately 2.19 gpp values using 16 quantization factors. Compared to the PSNR value, it has gained 5.03db more PSNR than Jpeg. In the case of the gpp value, it requires around 0.52 further gpp value than Jpeg. It has improved PSNR by sacrificing a smaller amount of gpp value. For Q=8, Jpeg achieved nearly 31.54db PSNR and 1.28 bpp value. Conversely, PALQA produces 36.76db PSNR and 1.69 gpp using 32 quantization factor. It is seen that PALQA has acquired 5.22db higher PSNR than Jpeg. At the same time, an additional around 0.41 gpp value is needed to gain those PSNR values. It has shown that further gpp is reasonable for attaining those PSNR values. For the Q=16 quantization factor, Jpeg obtained a 26.12db PSNR and 0.73 bpp value. On the other hand, PALQA achieves 31.16db PSNR and 0.94 gpp values using the Q=90 quantization factor. Compared to the gpp value, PALQA needs nearly 0.21 higher gpp than the Jpeg approach. At the time, in the case of PSNR, it earns around 5.04db, a further PSNR value than Jpeg. For Q=32, Jpeg acquires 22.39db PSNR and 0.32 bpp value, whereas PALQA achieves 30.56db PSNR and 0.84 gpp. The comparison result shows that PALQA provides an 8.2db higher PSNR value, but it also requires an extra 0.52 gpp. 

\begin{figure*}
\centering
    \subfigure [cameraman image]
    {
        \includegraphics[width=0.45\textwidth, height=5cm]{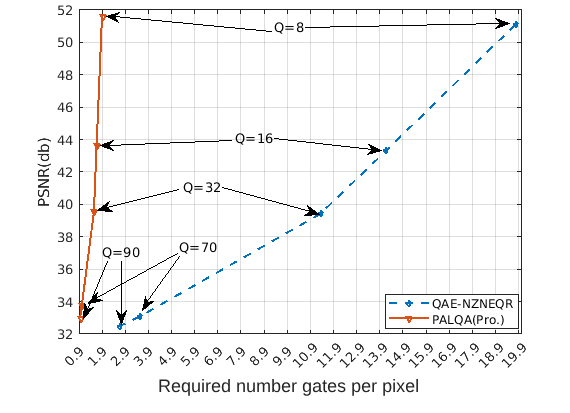}
        \label{gray_channel_cameraman_prop}
    }
    \subfigure[baboon image]
    {
        \includegraphics[width=0.45\textwidth, height=5cm ]{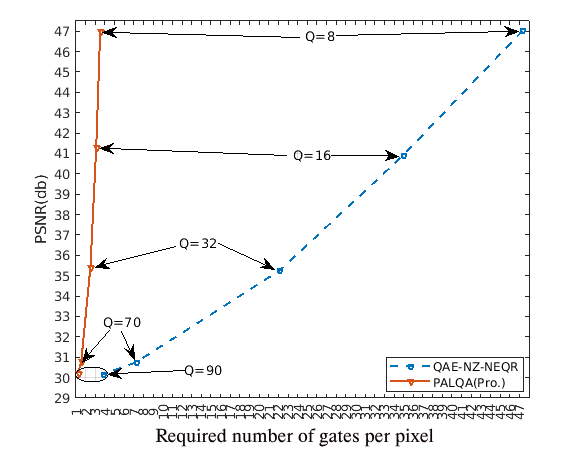}
        \label{gray__channel_baboon_prop}
    }
    \subfigure[scenery image]
    {
        \includegraphics[width=0.45\textwidth,height=5cm ]{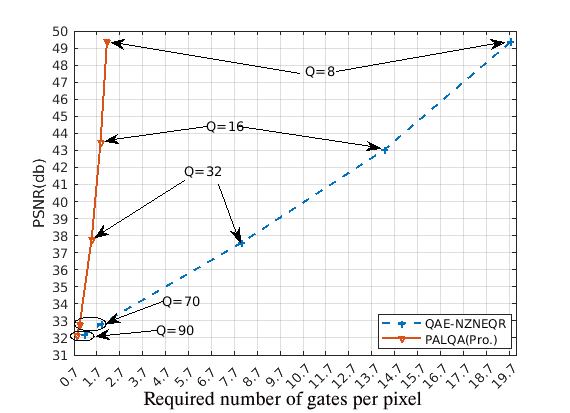}
        \label{gray__channel_scenery_prop}
    }
    \subfigure[pepper image]
    {
        \includegraphics[width=0.45\textwidth,height=5cm ]{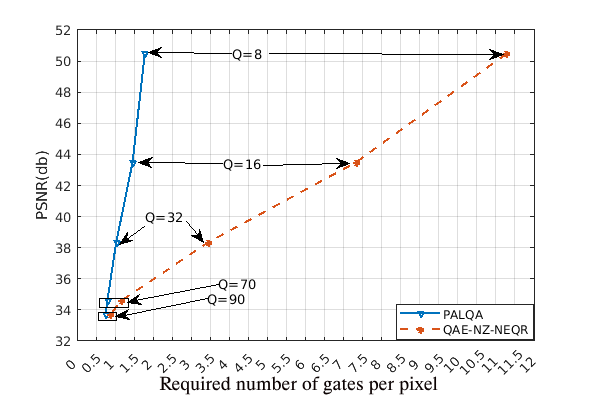}
        \label{gray_channel_peppers_prop}
    }
    \subfigure[airport image]
    {
        \includegraphics[width=0.45\textwidth,height=5cm ]{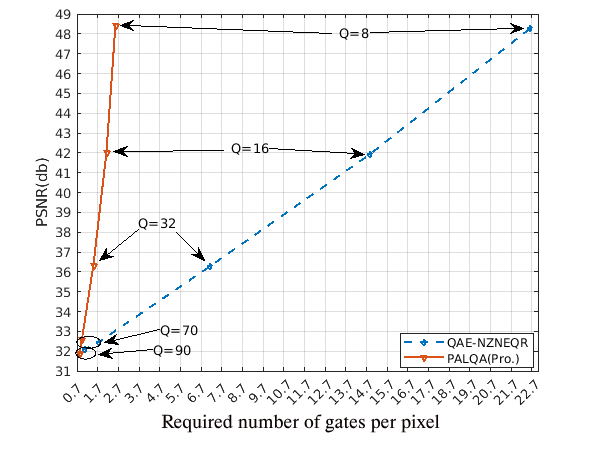}
        \label{gray_channel_airport_prop}
    }
     \subfigure[building image]
    {
        \includegraphics[width=0.45\textwidth,height=5cm ]{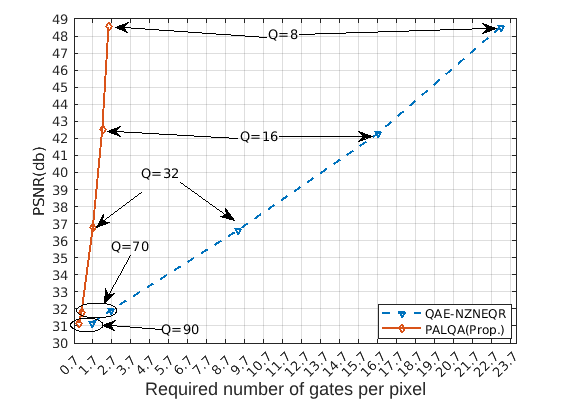}
        \label{gray_channel_building_prop}
    }
\caption{RD performance of the proposed approach compared to the NZ-NEQR based QAE approach for the gray channel of benchmark images, where qubit number 4 (LSB) of the X-position is considered as swap(trash) qubit.}
\label{Pro_appro_Traditional}
\end{figure*}

Figure~\ref{Pro_appro_Traditional} shows the computational result of the proposed approach compared to the NZ-NEQR-based quantum autoencoder approach using 8,16,32, 70, and 90 quantization factors for the considered images. Moreover, the comparison result of the cameraman image of the proposed approach compared to the NZ-NEQR-based QAE approach is depicted in Figure~\ref{gray_channel_cameraman_prop} using 8,16,32,70 and 90 quantization factor via lossy preparation approach. Over each quantization factor, the PALQA approach performs more robustly and is superior for both PSNR and gpp value than the NZ-NEQR-based QAE approach.  \par  

Figure~\ref{gray__channel_baboon_prop} depicts the RD performance result of the proposed approach compared to the NZ-NEQR-based QAE approach using 8,16,32,70, and 90 quantization factors for the baboon image. It has shown that by increasing the quantization value. In both approaches, RD performance value decreases. The RD result indicates that PALQA performs superior to NZ-NEQR-based QAE. Increasing the quantization factor makes the RD region's difference smaller. Figure~\ref{gray__channel_scenery_prop} demonstrates the RD performance of the  PALQA approach of the scenery image using 8,16,32,70, and 90 quantization factors via several classical and quantum-pre-processing-approaches-compared to NZ-NEQR-based QAE. The comparison result shows that for each quantization factor, PALQA requires less gpp but the same PSNR value as the NZ-NEQR-based QAE approach. It shows the RD performance of the PALQA approach using 8,16,32,70, and 90 quantization factors. \par
Figure~\ref{gray_channel_peppers_prop} depicts the gray channel of the pepper image using different quantization factors. A higher difference in gpp value is found for lower quantization factors, such as 8 and 16, but the same PSNR value is drawn for both approaches. A higher gpp difference is still found for a medium quantization factor, such as Q=32, but both approaches provide the same PSNR value. For the higher quantization factors such as Q=70 and 90, PALQA requires less gpp value but produces the same PSNR signal compared to the NZ-NEQR-based QAE approach. Overall, for every quantization factor, PALQA performs superior to the NZ-NEQR-based QAE approach. \par 

The proposed approach RD performance metrics for the airport image using 8,16,32,70, and 90 quantization factors via several pre-processing approaches is illustrated in Figure ~\ref{gray_channel_airport_prop}  compared to the NZ-NEQR based QAE approach. For Q=8, it provides 2.56 gpp and 48.43db PSNR values, while NZ-NEQR-based QAE draws 48.25db PSNR and 22.6 gpp values. The comparison result shows that gpp is the main fact that makes the difference in RD performance value. It requires 2.56 gpp to transmit a 48.43db PSNR signal, while NZEQR requires 22.6gpp to transmit a 48.25db signal. For Q=16,it requires approximately 2.12 gpp to transmit a 42.01db PSNR signal.\par On the other hand, NZ-NEQR-based QAE delivers a 41.92db PSNR value while it needs to transmit a 14.86 gpp value using the 16 quantization factor. The comparison result shows that PALQA requires 12.74 less gpp than the NZ-NEQR-based QAE approach. In the case of PSNR comparison, PALQA also gains 0.09db PSNR value compared to NZ-NEQR-based QAE. For Q=32, PALQA achieves 36.33db PSNR by transmitting around 1.49 gpp value. On the contrary, NZ-NEQR-based QAE produces 36.27db PSNR and 7.13 gpp values. Regarding the gpp comparison, PALQA needs only approximately 5.64 gpp less than the NZ-NEQR QAE approach. Moreover, PLAQA generates nearly 0.06db higher gpp than the NZ-NEQR QAE approach. For Q=70, PALQA provides almost 32.57db PSNR, while NZ-NEQR QAE produces 32.43db. Regarding the PNSR difference, PALQA generates roughly 0.14db higher PSNR than the NZ-NEQR QAE approach. At the same time, it also needs an additional 0.79 gpp value compared to the NZ-NEQR-based QAE approach. PALQA delivers 31.86db PSNR and 0.82 gpp using the Q=90 quantization factor. On the other hand, using a 90 quantization factor, the based QAE approach achieved about 32.09db PSNR by transmitting approximately 1.08 gpp. The comparison result shows that PALQA draws approximately 0.76db higher PSNR signal, including around 0.99 gpp less than the NZ-NEQR QAE approach. \par

Figure~\ref{gray_channel_building_prop} shows the required number of gates per pixel and PSNR value of the proposed approach compared with the traditional QAE-NZMEQR approach for the building image using 8, 16, 32, 70, and 90 quantization factors. For each quantization factor, the required number of gates per pixel (gpp) draws the noticeable difference. As the quantization factor increases, the gpp values decrease, and the image quality decreases. It is concluded that the proposed approach provides robust results for building images.  

\section{conclusion}
\label{CC}
This paper presents a novel PALQA quantum autoencoder for image compression, addressing compression issues in end-to-end communication systems. Unlike existing approaches, our strategy utilizes a transfer coefficient-based lossy quantum autoencoder with LSB (least significant bit) positional-aware qubit connections to tackle state connection compression problems.
Real-world image mapping poses a substantial barrier to quantum autoencoders because traditional mapping approaches cannot encode each pixel while considering its state connection with a quantum photonic circuit. Additionally, current autoencoder techniques rely on probabilistic results, and the mechanisms of state compression need clarification. Previous methods have also not thoroughly investigated compression capacity concerning quantum resource reconstruction.\par
Our suggested framework allows the creation of near-term quantum computer applications utilizing quantum-classical autoencoders. The transmitted signal is reconstructed through the use of a decoder. \\
We compared our method's compression ability to JPEG and NZ-NEQR-based quantum autoencoder techniques using the rate-distortion (RD) performance parameter. Due to the block-wise division of quantum transfer coefficients, LSB-based position encoding, and a lossy quantization procedure, our technique operates robustly and efficiently, as the findings demonstrate. The JPEG method is not appropriate for quantum compression since it lacks position encoding. Exploring additional enhancements to the PALQA circuit and its applications in other quantum computing fields can be future research.

\section*{Acknowledgment}
The author declares that there is no conflict of interest.

\bibliographystyle{unsrt}  


\end{document}